\documentclass[preprint,12pt, authoryear]{elsarticle}

\usepackage{amssymb}
\usepackage{amsmath}
\usepackage{xspace}
\usepackage{adjustbox}
\usepackage{float}

\journal{Journal}

\begin{document}

\begin{frontmatter}



\title{CBIDR: A novel method  for information retrieval combining image and data by means of TOPSIS applied to medical diagnosis}


\author[labcinaddress,ppgiaddress]{Humberto Giuri}\corref{mycorrespondingauthor}
\cortext[mycorrespondingauthor]{Corresponding author}
\ead{krohling.renato@gmail.com}

\author[labcinaddress,ppgiaddress]{Renato A. Krohling}
\ead{humbertogiuri@gmail.com}

\address[labcinaddress]{Labcin - Nature-inspired computing Lab.  - Federal University of Espírito Santo - Vitória, ES, Brazil}
\address[ppgiaddress]{Graduate Program in Computer Science, PPGI, UFES - Federal University of Espírito Santo - Vitória, ES, Brazil}

\begin{abstract}
 Content-Based Image Retrieval (CBIR) have shown promising results in the field of medical diagnosis, which aims to provide support to medical professionals (doctor or pathologist). However, the ultimate decision regarding the diagnosis is made by the medical professional, drawing upon their accumulated experience. In this context, we believe that artificial intelligence can play a pivotal role in addressing the challenges in medical diagnosis not by making the final decision but by assisting in the diagnosis process with the most relevant information. The CBIR methods use similarity metrics to compare feature vectors generated from images using Convolutional Neural Networks (CNNs). In addition to the information contained in medical images, clinical data about the patient is often available and is also relevant in the final decision-making process by medical professionals. In this paper, we propose a novel method named  CBIDR, which leverage both medical images and clinical data of  patient, combining them through the ranking algorithm TOPSIS. The goal is to aid medical professionals in their final diagnosis by retrieving images and clinical data of patient that are most similar to query data from the database. As a case study, we illustrate our CBIDR  for diagnostic of oral cancer including histopathological images and  clinical data of patient. Experimental results in terms of accuracy achieved 97.44\% in \textit{Top-1} and 100\% in \textit{Top-5} showing the effectiveness of the proposed approach.

\end{abstract}




\end{frontmatter}


\section{Introduction}
In recent years, Convolutional Neural Networks (CNN) for image classification  have provided promising results \citep{cnn-classification}, \citep{cnn-image-recognition}, \citep{Wang2021}, \citep{Nawaz2022}, among others. Applications focused on the medical field \citep{Yang2021} have also been explored achieving good results such as the classification of skin lesion images \citep{Wu2022}, classification of histopathological images of oral cancer using different neural network architectures \citep{Bea2023}, among others. 
 \cite{diagnostic2019} used 7 different types of CNNs to classify metastases in breast cancer histopathological images, achieving results comparable to 11 experienced pathologists. However, many challenges still need to be addressed when using CNNs for medical image classification. The lack of publicly labeled databases by experts, class imbalance, and the high resolution of histopathological images are some of the challenges faced and opportunities that arise in this area of study \citep{Tizhoosh2018}, \citep{Goyal2020}.

Image retrieval aims to provide support to medical professionals during diagnosis. However, the ultimate decision regarding the diagnosis is made by the doctor or pathologist, drawing upon their accumulated experience. In this context, we believe that deep learning can play an important role in addressing the challenges in medical diagnosis—not by making the final decision but by assisting in the diagnosis process by suggesting the most relevant information to the doctors. To overcome the challenges while leveraging the expertise of pathologists, applications and frameworks designed to assist pathologists are starting to emerge, showing promising results. Advances have been made in both accuracy and speed when applied to areas such as Image Retrieval \citep{Kuo2016}, especially for applications that rely on the visual characteristics of the image. This field of study is called Content-Based Image Retrieval (CBIR).  \cite{Barhoumi2021} propose a tool to retrieve similar images related to an input image by measuring the distance to the feature map generated by the convolution and pooling processes of a CNN.

There is an extensive body of literature in the field of image retrieval, with the majority of works focusing on content-based image retrieval using neural networks as feature extractors. For example, \cite{Alzubi2017} achieved an accuracy of nearly 95.7\% on the Oxford 5K database \citep{Philbin2007} and 88.6\% on the Oxford 105K database \citep{Philbin2007} using two parallel CNNs as feature extractors.

When applied to the medical field, CBIR systems have proven to be a valuable tool in diagnostic support. For instance, \cite{Choe2022} improved the diagnostic accuracy of interstitial lung disease and the agreement among experts with varying levels of experience using CBIR for chest computed tomography images with deep learning.

\cite{Komura2018} developed a web application called Luigi, which retrieves histopathological images similar to various types of cancer using a pretrained convolutional neural network to extract features. \cite{Hegde2019} proposed the SMILY framework (Similar Medical Images Like Yours). SMILY is a tool for searching for similar histopathological images trained without the use of labeled histopathological images in training. SMILY was trained and tested on The Cancer Genome Atlas Program (TCGA) database, achieving high accuracy and efficiency in image retrieval. The tool achieved an average precision and recall of 96\% and 98\%, respectively. It was able to search a database containing tens of thousands of samples in less than 1 second, outperforming various state-of-the-art algorithms in both accuracy and time efficiency. Due to the high dimensionality of this type of image, as the number of captured and usable images in the system increases, image retrieval operations become increasingly computationally expensive. \cite{chen2022fast}, in addition to discussing this point, they propose a fast and scalable image retrieval system for slides, which are glass slides containing samples of biological tissue prepared for microscopic analysis, achieved through deep self-supervised learning. The algorithm, named SISH (Self-Supervised Image Search for Histology), achieved nearly constant retrieval time regardless of the image database size.

In order to assist pathologists in their decision-making process regarding the diagnosis, \cite{Belga} proposed a distributed approach to histopathological image retrieval to handle millions or even billions of images. This approach utilizes convolutional neural network architectures as feature extractors and the Faiss library \citep{Faiss} for fast approximate nearest neighbor search, which is commonly used in image retrieval systems. \cite{Hashimoto2023} introduced a new method for similar image retrieval in histopathological images stained with hematoxylin and eosin for malignant lymphoma. When using a whole-slide image as the input query, which can be of high dimensionality, it is desirable to retrieve similar cases while focusing on areas of pathological importance, such as tumor cells. 

In an effort to incorporate the information contained in clinical data into computer-aided diagnosis (CAD), \cite{deLima2023} assessed the importance of supplementary data in the analysis of histopathological images of oral leukoplakia and carcinoma. The study concluded that clinical and demographic data positively influenced the accuracy of the models, resulting in a 30\% improvement in balanced accuracy.

In the majority of cases, image retrieval work relies on content-based approaches using convolutional neural networks. In cases of  histopathological images, these images are often collected along with clinical data of patient. This information is quite rich and has significant potential for providing insights into the disease. Therefore, this study will investigate the combination of images and clinical data in the process of assisting pathologists to achieve a more accurate diagnosis.

In this work, a novel approach for image retrieval is proposed combining convolutional neural networks used for image feature extraction with clinical data of patient. As a case study, a dataset NDB-UFES consisting of histopathological images of oral cancer and patient data \citep{falcaosab}, is used. This dataset contains three classes considered for diagnosis: oral cavity squamous cell carcinoma, leukoplakia with dysplasia, and leukoplakia without dysplasia. Additionally, clinical and sociodemographic patient information is included, such as gender, lesion location, smoking habits, alcohol consumption, age, and sun exposure \citep{falcaosab}.

The main contributions of this work are twofold:
\begin{itemize}
\item we propose for the first time, as far as we know, a new method for Content Based Image and Data Retrieval (CBIDR) by means of the ranking algorithm TOPSIS (Technique for Order Preference by Similarity to Ideal Solution). 
\item we apply the method CBIDR to a case study involving diagnostic of oral cancer using NDB-UFES dataset, which consists of histopathological images and clinical data of patient in order to illustrate the approach and show its feasibility.
\end{itemize}

In the remaining sections of the paper, we follow this structure: Section 2 briefly describes Convolutional Neural Networks, which are used in this work as feature extractors. Section 3  explains the basic concepts of Content-Based Image Retrieval (CBIR) and proposes a new methodology (CBIDR) to combine images and data through the ranking algorithm TOPSIS. Section 4 presents and discuss the obtained results and Section 5 ends up the paper with conclusions as well as directions for future work.

\section{Convolutional Neural Networks}
\subsection{Background}

\begin{figure}[!htb]
\centering
\includegraphics[width=1\textwidth]{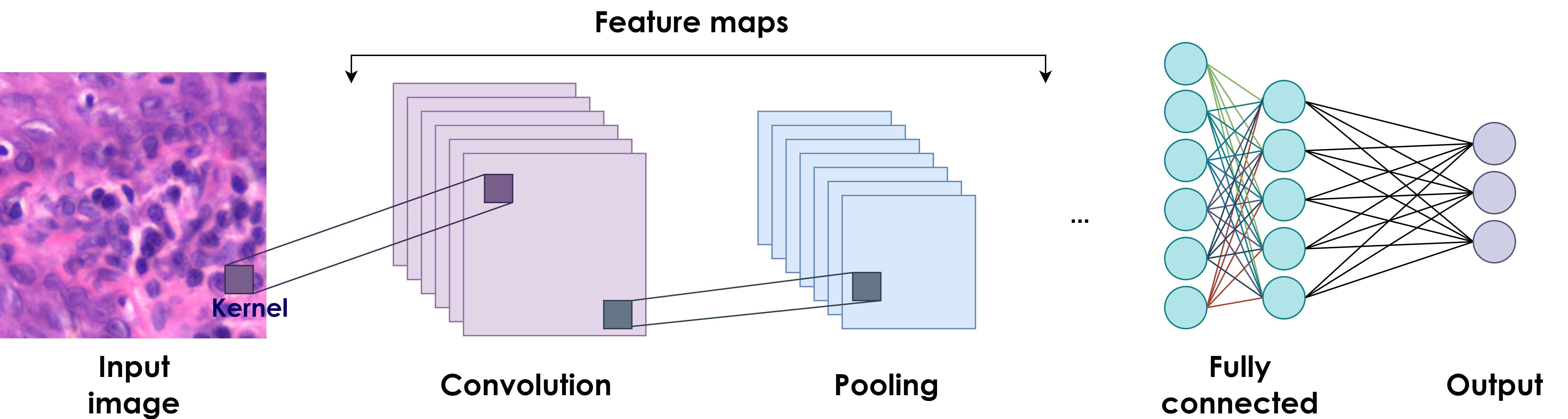} 
\caption{Illustration of an architecture of convolutional neural network with convolutional layers, pooling, and fully connected layers.}
\label{fig2_color}
\end{figure}

Convolutional Neural Networks (CNNs), proposed by \cite{LeCun1989}, were originally used for handwritten postal code recognition. However, they have found applications in various fields, especially in dealing with high-dimensional data. \cite{Krizhevsky2017} demonstrated the potential of CNNs by winning the "ImageNet Large Scale Visual Recognition Challenge" competition \citep{ILSVRC15} using a multi-layered CNN known as AlexNet. This result showcased the viability and versatility of CNNs. CNNs architecture consist of three types of layers: convolutional, pooling, and fully connected layers \citep{Goodfellow-et-al-2016}. Figure ~\ref{fig2_color} illustrates a CNN  architecture.

The convolutional layer consists of filters, often referred to as kernels. These trainable filters traverse the entire input data, applying the convolution process. In most cases, the input data is an image. These input filters are trainable, meaning they seek to learn the best configuration so that the output of this layer is a feature map that effectively describes the input. Typically, towards the end of the convolutional layer, an activation function is applied to the resulting feature map. This activation function introduces non-linearity to the model, and enhances important features, allowing the CNN to learn and model complex relationships within the input data. One commonly used activation function in CNNs is the Rectified Linear Unit (ReLU) \citep{Goodfellow-et-al-2016}. It is often chosen for its simplicity and effectiveness, helping to identify activation regions in the feature maps of the convolutional layer.

The pooling operation aims to reduce the dimensionality of the feature map resulting from the convolutional layer \citep{Goodfellow-et-al-2016} by selecting the most important features.  During the pooling operation, a sliding window, for example, 2x2, is applied across the output of the previous layer. This window's purpose is to summarize the information within that area into a single value, thus achieving dimensionality reduction and highlighting relevant features. This technique helps simplify and compress information, enabling more efficient processing and a more compact representation of the data. The two most well-known pooling techniques are Max Pooling (MaxPool) and Average Pooling (AvgPool).

The final part of a CNN,  consists of a feedforward neural network, also known as a Multilayer Perceptron (MLP) \citep{Goodfellow-et-al-2016}. Typically, it acts as the classification block in the CNN, taking the feature map as input and producing an output for a given input. A feedforward neural network consists of three main types of layers: the input layer, hidden layers, and the output layer \citep{Goodfellow-et-al-2016}. The input layer, as the name suggests, receives the input data. In a CNN, this input data represents the feature map. The hidden layers, situated between the input and output layers, are responsible for learning and storing abstract representations of the input data. In these hidden layers, each neuron receives outputs from neurons in the previous layer, performs calculations based on its weights and activation function, and passes the result to the next layer. Ultimately, in the output layer, the model's output is computed.

In convolutional neural networks, before feeding the fully connected (FC) layer with the resulting feature map, it is necessary to perform the flatten operation. Typically, this feature map is a three-dimensional vector, with depth representing the number of filters used. The FC layer needs to receive a one-dimensional vector, so the feature map is flattened into a one-dimensional vector, allowing the feedforward neural network to classify the input data.

Most convolutional neural networks, such as ResNet \citep{resnet}, MobileNetV2 \citep{mobilenetv2}, and DenseNet-121 \citep{densenet}, sequentially employ blocks of convolution and pooling layers multiple times. This architecture allows the model to capture information at different scales. Since convolutional neural networks can be used for dimensionality reduction of input data, providing a one-dimensional vector, removing the fully connected layers (FC) from the network yields the feature vector. This task is known as feature extraction \citep{ATASEVER2023} and involves representing the original input data in a compact form without losing information quality, capable of capturing essential features while discarding deemed irrelevant information. Feature extraction enables high-dimensional data, such as images, to have their key characteristics mapped to reduced dimensionality, improving efficiency. 

\subsection{Training Convolutional Neural Networks}
For training a neural network, the backpropagation algorithm, as introduced by \cite{Rumelhart1986}, is commonly used. The goal of this algorithm is to adjust the network's weights to minimize the associated cost function. Initially, the network's weights and bias are randomly initialized. The network is then fed with the training data, and the output of each neuron is calculated until reaching the final layer. After calculating the final network outputs, the error can be computed by comparing the obtained output with the expected output. Next, the error is propagated backward through the network, determining the contribution of error from each neuron. This is accomplished using gradient descent, which involves computing the partial derivative of the error with respect to the synaptic weights of each connection. The weights are updated for each connection by an amount proportional to the gradient descent, multiplied by a learning rate. This update is aimed at reducing the error. These steps are repeated for all training data, constituting one epoch. The training of a neural network typically involves multiple epochs, often predetermined, or until convergence is achieved. In summary, backpropagation is a key algorithm for training neural networks, and it iteratively adjusts the weights to minimize the error between predicted and actual outputs during training. 

The goal of training a neural network is to find the weights and bias configurations for a given problem, and for a CNN, this includes finding the optimal filter configurations for the convolutional layers. In the training process of a convolutional neural network, labeled data must be provided to enable the network to make predictions. By using a loss function \citep{Goodfellow-et-al-2016}, one can measure the comparison between predicted classes and true classes. The objective is to minimize this resulting error to obtain the correct predictions. The Cross-Entropy loss function is commonly used and it is described by:

\begin{equation}
\label{eq-cross}
\operatorname{Loss}(x, y)=-\sum_i^N x_i \log y_i
\end{equation}

\noindent where $x_i$ represents the true probability of image $i$ belonging to the correct class, $y_i$ is the predicted probability by the neural network that image $i$ belongs to the correct class and $N$ is the total number of training examples.

In feature extraction tasks, loss functions based on ranking are commonly used. They aim to predict relative distances between inputs. One such loss function is the Margin Loss \citep{margin}, defined as:

\begin{equation}
\mathcal{L}_{\text {margin }}=\sum_{(i, j) \in P} \gamma+\mathbb{I}_{y_i=y_j}\left(d\left(\phi_i, \phi_j\right)-\beta\right)-\mathbb{I}_{y_i \neq y_j}\left(d\left(\phi_i, \phi_j\right)-\beta\right)
\end{equation}

\noindent where $\beta$ is an adjustable parameter, regulated during training.
The goal here is to learn a function $\phi$ such that $d_{\phi}(x_a, x_n) - d_{\phi}(x_a, x_p) < \gamma$, assuming $x_a$ is an anchor image from a specific class, $x_n$ is an image from a different class than $x_a$, and $x_p$ is an image from the same class as $x_a$. Thus, this loss function encourages images of the same class to be closer while pushing images of different classes farther apart.

To minimize the training error, optimization algorithms like Stochastic Gradient Descent (SGD) \citep{ruder2016overview} and Adam \citep{kingma2017adam} are commonly used. This class of algorithms aims to minimize the loss function to its optimal value by updating the weights and biases of the network. In real-world applications, convolutional neural networks often leverage transfer learning. So, a pre-trained CNN, trained on a large dataset like ImageNet \citep{Deng2009}, is used. This allows to save the previously learned network parameters and fine-tune them on your specific target dataset.

\section{Image Retrieval}

\subsection{Preliminaries}

The main methods for image retrieval \citep{MRiad2012}  are text-based image retrieval (TBIR), content-based image retrieval (CBIR), and hybrid approaches.  In TBIR, the text associated with the image is considered to determine its content, making it possible to retrieve similar elements based on the text present. In the case of CBIR, features such as color, shape, and texture are used to index and compare the similarity of input images with those in the database. The third method, a hybrid retrieval, aims to combine the two previously mentioned approaches to achieve better results.

To determine the similarity between the input image and those in the database, a similarity metric is computed based on the extracted features \citep{MRiad2012}. Therefore, research in this area focuses on how to extract features that best describe the images, aiming for a detailed comparison. With the advancement of CNNs for feature extraction tasks, they have been introduced into CBIR systems for this purpose. For instance, \cite{Rian2019} investigated the use of convolutional neural networks as feature extractors for image retrieval in the iNaturalist database \citep{INaturalist}, achieving accuracy of 89\%.

The main advantage of using CNNs for image retrieval lies in their ability to extract features with a richness of information that describes the image. \cite{Gkelios2021} conclude that when CNNs are employed for feature extraction (or descriptors), they produce descriptors with high discriminative power in a computationally efficient manner. These feature vectors enable the comparison of two images using similarity metrics like the Euclidean distance, which quantifies how similar two images are.

\cite{Datta2008} address the challenges and trends in the field of image retrieval, presenting and reviewing the main similarity metrics used in this area. Consider two feature vectors, $a$ and $b$. The Euclidean distance between them is defined as:

\begin{equation}
D(a, b)=\sqrt{\sum_{i=1}^p\left(a_{i} - b_i\right)^2}
\label{eq-l2}
\end{equation}

\noindent where the sum of the square root of the difference between $a$ and $b$ in their respective dimensions. There are several metrics that can be used to calculate similarity between feature vectors but this is not the focus of this work and may be explored elsewhere.

\subsection{Standard CBIR Method}

\begin{figure}[!htb]
\centering
\includegraphics[width=1.0\textwidth]{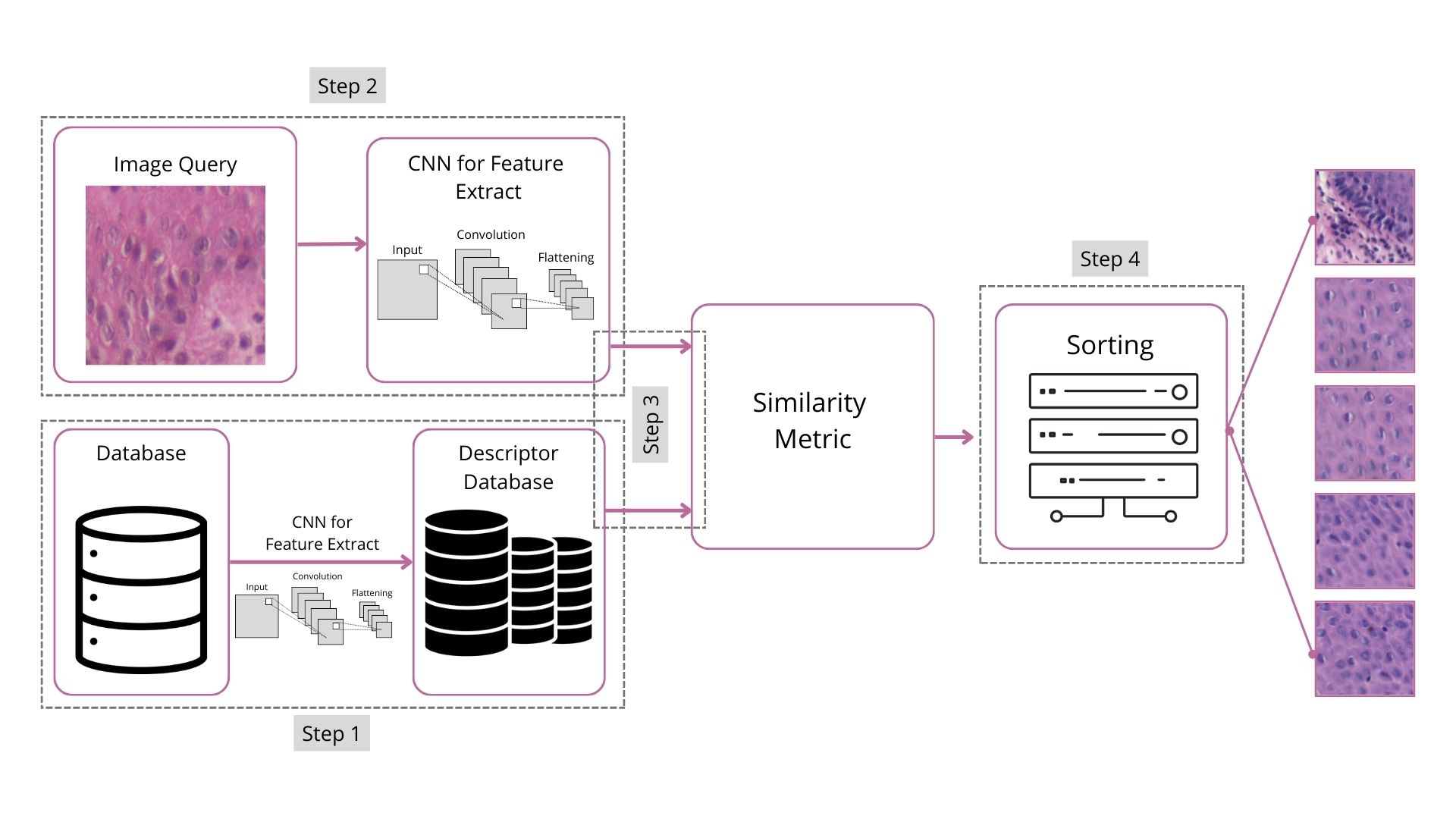} 
\caption{Illustration of an image retrieval process using only images. The first step represents the creation of the descriptor database. In this step, the images in the database are passed through a CNN to extract features. Step 2 describes the process of feature extraction from the query image. In step 3, the feature vector resulting from step 2 is compared to those stored in the descriptor database, resulting in similarities between the images. Step 4 concludes the process by sorting in ascending order, and displaying the first images being the most similar.}
\label{fig-image-retrieval}
\end{figure}

The image retrieval task is divided into 4 steps. The first one is to extract the feature vectors from all images in the database using a CNN as a feature extraction mechanism to create the descriptor database. The second step is to receive a query image that will serve as the input image for the search. After extracting the feature vector from the query image using the same CNN as in the first step, distances are calculated between the input image and the descriptors in the database. This allows ranking the compared descriptors based on the metric and selecting the top images with the highest similarity value. Figure~\ref{fig-image-retrieval} illustrates the CBIR method using CNN.

The Faiss \footnote{https://faiss.ai/} library  \citep{Faiss} is used to store the descriptor database. It performs the actual search for the $k$ nearest neighbors through an exhaustive search. The distance $\left|\mathbf{x}_j-\mathbf{y}_i\right|^2$ is expanded to $\left|\mathbf{x}_j\right|^2+\left|\mathbf{y}_i\right|^2-2\left\langle\mathbf{x}_j, \mathbf{y}_i\right\rangle$, where $\langle.,$.$ \rangle$ is the dot product. The first two terms are precomputed, so the bottleneck lies in evaluating $\left\langle\mathbf{x}_j, \mathbf{y}_i\right\rangle$.

%%% Início de seção. %%%
\subsection{Method Combining Image and Data}
\label{sec-fundteo-secoes3}

As mentioned earlier, an individual's lifestyle is directly related to the occurrence of various diseases. Therefore, clinical and sociodemographic data, such as smoking habits, alcohol consumption, gender, and others, are usually collected along with the images and provide relevant information for diagnosis. Since it is possible to numerically define how similar two images are by extracting their respective feature vectors and measuring, using a similarity metric, the distance between them, it is also possible to quantify how close two sets of clinical information are. By organizing the information into a binary vector, where 0 corresponds to the absence of a particular feature - such as whether or not alcohol is consumed - and 1 corresponds to its presence, it is possible to measure the distance between two binary vectors using distance metrics.

One of the most commonly used metrics to calculate similarity between two binary vectors is the Hamming distance. As highlighted by Bookstein (2002), the Hamming distance is commonly used to quantify the extent to which two bit strings of the same dimension differ. An early application was in the theory of error-correcting codes (Hamming, 1980), where the Hamming distance measured the error introduced by noise over a channel when a message, typically a sequence of bits, is transmitted between the source and destination. To calculate the Hamming distance, a logical XOR operation is performed between the positions of the two input vectors. The number of 1 bits in the resulting vector determines the value of the distance.

By using CNNs in the image retrieval process, distances between the extracted features of each image are ranked. By incorporating clinical data into the analysis, two distance metrics are considered: one for images and another for clinical data. Therefore, when performing an operation with image and data retrieval, it is possible to compare an input query image and data with those  image features and data stored in the database, resulting in two distances for each element in the database. Thus, the system will generate an Mx2 matrix, where each column represents one of the obtained distances. To identify the option most similar to the query image and data, it is necessary to rank an Mx2 matrix rather than a one-dimensional vector of size M. For this purpose, the TOPSIS algorithm for ranking is employed, which has been applied in different knowledge domains \citep{atopsis}.

\subsubsection{TOPSIS}

TOPSIS (Technique for Order Preference by Similarity to Ideal Solution)  proposed by \cite{Hwang1981} is a decision-making technique, which works by evaluating the performance of alternatives based on their similarity to the ideal solution. Multi-criteria decision-making problems are characterized by a decision matrix $D$, where each row represents an alternative and each column represents a criterion. This matrix is used to organize the relevant information for the decision problem, allowing for the comparison of alternatives with respect to the established criteria. Based on this matrix, the steps of the TOPSIS method are applied to determine the preferred alternative based on a combination of criteria and their weights. TOPSIS is a valuable technique for multi-criteria decision-making and is used in various fields to support decision processes when there are multiple  alternatives  and criteria to consider. Multi-criteria decision-making problems are characterized by a decision matrix  $D$:

\begin{equation}
\begin{array}{c}
\hspace{1.8cm} C_1 \hspace{0.4cm} \ldots \hspace{0.3cm} C_n \\
\begin{array}{ccc}
D=\begin{array}{ccc}
A_1 \\
\vdots \\
A_m
\end{array}\left(\begin{array}{ccc}
x_{11} & \ldots & x_{1 n} \\
\vdots & \ddots & \vdots \\
x_{m 1} & \cdots & x_{m n}
\end{array}\right) \\
\end{array}
\end{array}
\label{eq-d-matriz}
\end{equation}

\noindent where the rows of this matrix represent the alternatives ($A_1, A_2, ..., A_m$). The columns are the criteria analyzed in the problem ($C_1, C_2, ..., C_n$). The elements of the matrix, $x_{ij}$ determine the performance of alternative $A_i$ evaluated on criterion $C_j$. Each criterion has a weight that affects the decision-making process, defined by $W = (w_1, w_2, ..., w_n)$, where $w_j$ is the weight for criterion $C_j$. It is important to note that the sum of all weights must equal 1. There are two types of criteria: cost and benefit. The difference between them is that for the cost criterion, the focus is on minimizing, meaning that a lower value is better, while for the benefit criterion, it is exactly the opposite. In this work, both criteria are cost criteria, as we use the distances between images and between clinical data as the criteria, aiming for the smallest distance values possible.

Before using TOPSIS, it is necessary to normalize the decision matrix $D$, transforming it into the matrix $R = [r_{ij}]_{m \times n}$, allowing for comparisons between all criteria. To normalize the decision matrix, the following formula is applied:

\begin{equation}
r_{ij} = \frac{x_{ij}}{\sqrt {\sum_{i=1}^m x^2_{ij}}}
\label{eq-normalizar}
\end{equation}

\noindent Once normalized, the resulting matrix $R$ has its values weighted by the weight vector $W$, generating a weighted matrix $P = [p_{ij}]_{m \times n}$, calculated as follows:

\begin{equation}
p_{ij} = w_j \times r_{ij}
\label{eq-p-mult}
\end{equation}

\noindent Upon obtaining matrix $P$, the TOPSIS algorithm begins. First, the positive ideal solution ($A^+$) and the negative ideal solution ($A^-$), respectively, for the benefit and cost criteria are identified as follows:

\begin{equation}
\begin{split}
A^+ = (p^+_1, p^+_2, \hdots, p^+_n) \\
A^- = (p^-_1, p^-_2, \hdots, p^-_n)
\end{split}
\label{eq-A}
\end{equation}

\noindent with:

\begin{equation}
\begin{split}
p^+_j = 
\left\{\begin{matrix}
max_i(p_{ij})\\ 
min_i(p_{ij})
\end{matrix} \right. \\
p^-_j = 
\left\{\begin{matrix}
min_i(p_{ij})\\ 
max_i(p_{ij})
\end{matrix} \right.
\end{split}
\label{eq-pj}
\end{equation}

\noindent where the maximum operator is used for benefit criteria, and the minimum operator is used for cost criteria. This step aims to select the best performance for each criterion, whether it is the cost criterion (where the best performance is the lowest value) or the benefit criterion (where the best performance is the highest value). Next, two vectors, $A^+$ and $A^-$, are formed to represent the ideal performance. Then, for each alternative $A_i$, the Euclidean distance from each element to the positive solution vector $A^+$ and the negative solution vector $A^-$ is calculated as follows:

\begin{equation}
\begin{split}
d^+_i = \sqrt {\sum_{j=1}^n (d^+_{ij})^2} \\
d^-_i = \sqrt {\sum_{j=1}^n (d^-_{ij})^2}
\end{split}
\label{eq-distancias}
\end{equation}

\noindent where, $d^+_i = p^+_j - p_{ij}$ and $d^-_i = p^-_j - p_{ij}$. The next step is to determine the relative closeness \begin{math} \xi_i \end{math}
based on the obtained distances, calculated as follows:

\begin{equation}
\xi_i = \frac{d^-_i}{d^+_i + d^-_i}
\label{eq-xi}
\end{equation}

\noindent this allows the selection of the alternative closest to the positive ideal solution and farthest from the negative ideal solution.

Finally, the alternatives in the  \begin{math} \xi \end{math} vector are ranked, with the best solutions having higher values of  \begin{math} \xi_i \end{math}, indicating that they are closer to the ideal solution.

\subsubsection{The Proposed Method CBIDR }
\begin{figure}[!htb]
\centering
\includegraphics[width=1.0\textwidth]{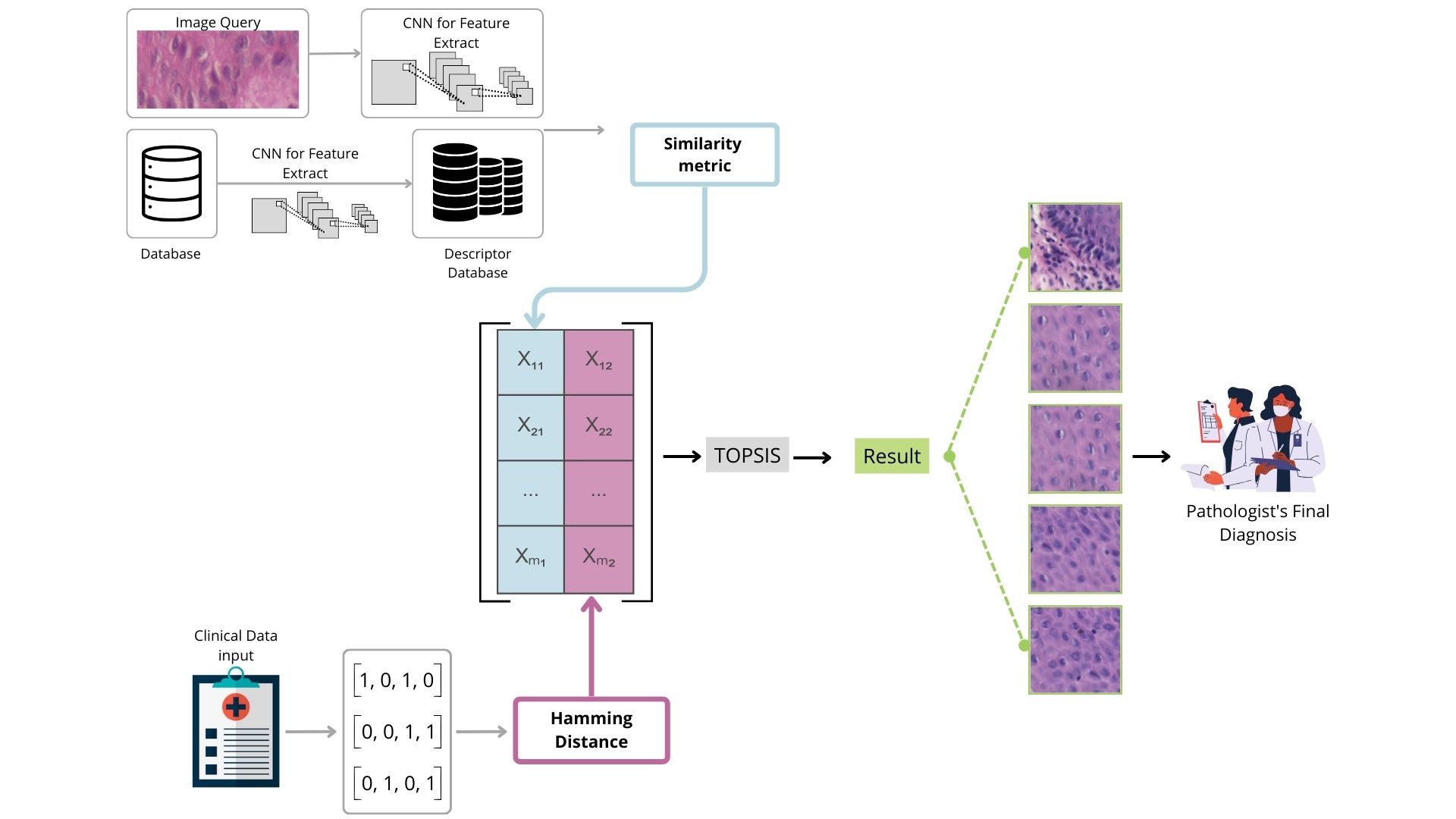} 
\caption{Illustration of the image retrieval process using CNN for feature extraction and clinical data using the TOPSIS technique. The two distance vectors are generated and passed as input to TOPSIS, producing ranked results that will be presented to the medical professionals for the final diagnostic.}
\label{fig-image-retrieval-full}
\end{figure}

The novel proposed Content-Based Image and Data Retrieval, for short, CBIDR is described in the following. The image retrieval process using convolutional neural networks combined with patient clinical data through the TOPSIS algorithm begins by creating a descriptor database using a CNN as the feature extraction mechanism. Next, an input image is passed to the system, and its features are extracted using the same CNN as in the previous step. Subsequently, the distance between the input feature vector and the descriptor database is measured, generating the first vector $X_1$, with a size of $m$, where $m$ is the number of descriptors present in the database. Continuing, the clinical information linked to the input image is compared, utilizing the Hamming distance, with the information in the database, generating the second distance vector $X_2$, also of size $m$. Combining the two distance vectors results in a decision matrix $D$ with dimensions $m$ x 2. This matrix serves as input for the TOPSIS technique, which, along with its weights and criteria, calculates and ranks the best alternatives. The complete process of image retrieval using convolutional neural networks combined with  clinical data of patient and the TOPSIS algorithm is shown in Figura~\ref{fig-image-retrieval-full}.

\section{Experimental Results}
\label{sec-projeto}

In this section, the results obtained for retrieval of histopathological images of oral cancer, which combines convolutional neural networks (CNNs) to extract image features with clinical data of patient, is presented. Three CNNs were employed to evaluate the quality of feature extraction. Firstly, the dataset used in the experiments is introduced. Next, the configurations used in the experiments are presented and discussed. Finally, the results are discussed.

\subsection{Dataset}
\label{sec-projeto-dataset}

\begin{figure}[!htb]
\centering
\includegraphics[width=0.8\textwidth]{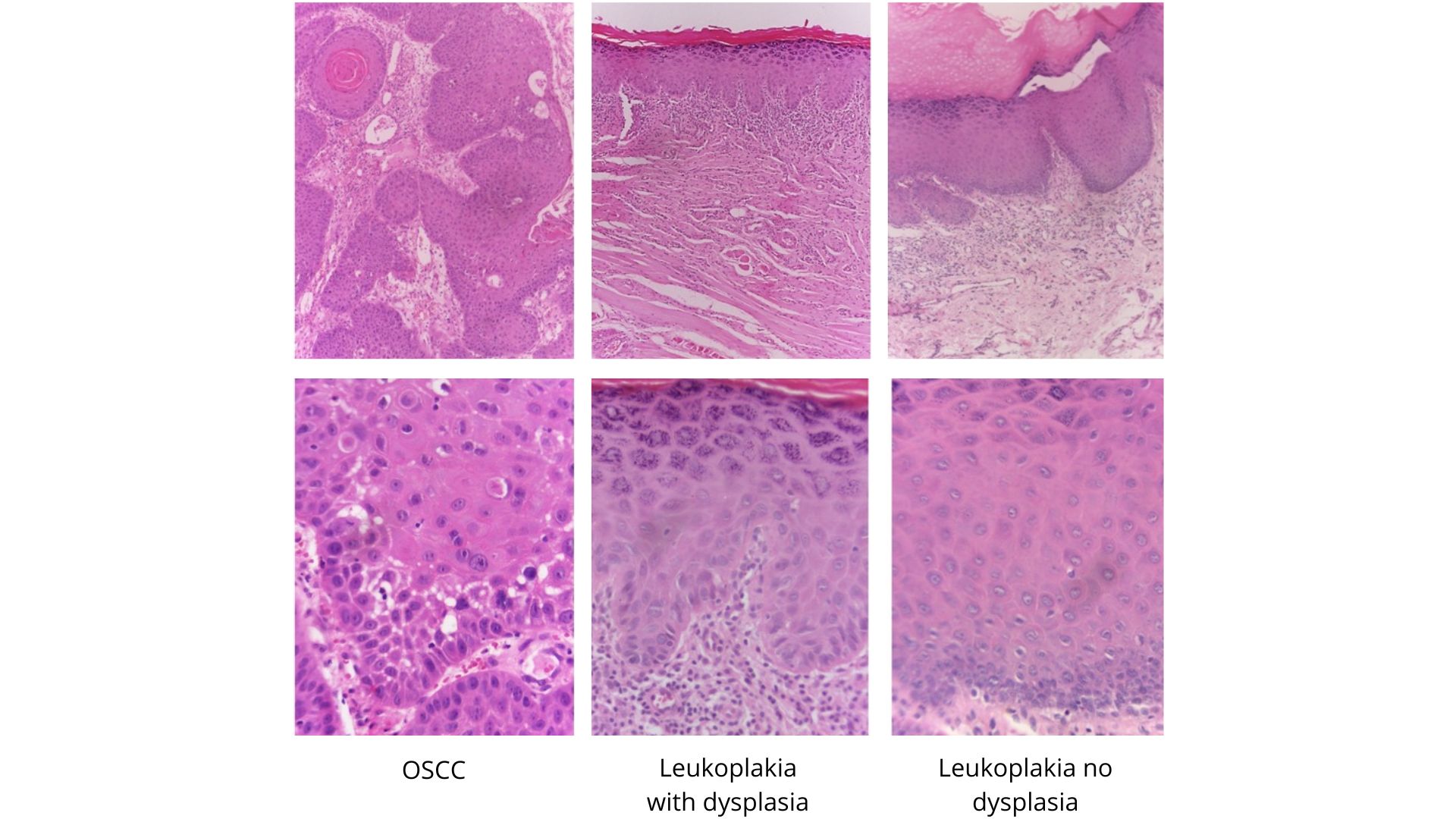} 
\caption{Examples of histopathological images used in this study.}
\label{fig-exemplos-intro}
\end{figure}

In this work, the dataset NDB-UFES \citep{falcaosab}  resulted from an extension project carried out at the Federal University of Espírito Santo, collected histopathological images of patients diagnosed with oral cavity squamous cell carcinoma and oral leukoplakia between January 2010 and December 2021. Figure ~\ref{fig-exemplos-intro} shows an example of histopathological images for each class in the NDB-UFES dataset. It contains originally, 237 images (2048 x 1536 pixels) and clinical and socio-demographic data were also collected from the respective patients, including gender, lesion location, cigarette usage, alcohol consumption, age, and sun exposure \citep{falcaosab}.

Next, the \textit{PatchExtractor} from the Python library \textit{Scikit-Learn} \citep{scikit-learn} was used to generate patches for each image, increasing the dataset to 3753 images (512 x 512 pixels). These 3753 images were labeled by pathologists, resulting in a database called P-NDB-UFES, which contains 1930 images (51.29\%) for the class leukoplakia with dysplasia, 707 images (18.79\%) for the class leukoplakia without dysplasia, and 1126 images (29.92\%) for the class oral squamous cell carcinoma (OSCC). 

\begin{figure}[!htb]
\centering
\includegraphics[width=0.8\textwidth]{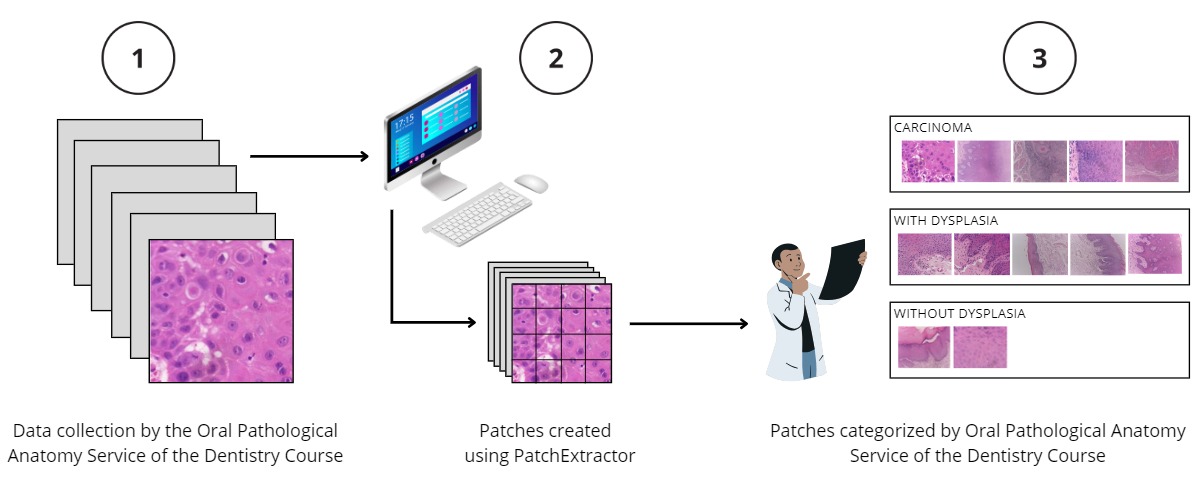} 
\caption{Generation process of histopathological patches images.}
\label{annotation}
\end{figure}

Table~\ref{tab-base-dados} presents the distribution of images by class in the dataset and Figure~\ref{annotation} shows examples of  histopathological patches images in the database.

% Tabela
\begin{table}[!htb]
    \centering
    \begin{adjustbox}{width=1\textwidth}
    \begin{tabular}{cccc}
    \hline
    \multicolumn{2}{c}{NDB-UFES} & \multicolumn{2}{c}{P-NDB-UFES} \\ \hline
    Class   & Number of images   & Class    & Number of images    \\ \hline
    \begin{tabular}[c]{@{}c@{}}Leucoplakia \\ with Dysplasia\end{tabular}  & 89 & \begin{tabular}[c]{@{}c@{}}Leucoplakia \\ with Dysplasia\end{tabular}    & 1930 \\ 
    \begin{tabular}[c]{@{}c@{}}Leucoplakia\\ without Dysplasia\end{tabular} & 57 & \begin{tabular}[c]{@{}c@{}}Leucoplakia \\ without Dysplasia\end{tabular} & 707  \\ 
    OSCC    & 91                 & OSCC     & 1126                \\ \hline
    \end{tabular}
    \end{adjustbox}
    \caption{Division of Images by Class in the NDB-UFES and P-NDB-UFES Databases.}
    \label{tab-base-dados}
\end{table}

\subsection{Experiments Setting}
\label{sec-projeto-conf}

The experiments conducted were divided into two parts: Experiment I and II. Experiment I represents the configuration using only image features (CBIR) as baseline, while Experiment II utilizes the TOPSIS algorithm to combine images and clinical data of patient.

In both experiments, three convolutional neural network architectures were used to perform the task of extracting image features. These architectures include ResNet50 \citep{resnet}, DenseNet-121 \citep{densenet} and MobileNetV2 \citep{mobilenetv2}. Table~\ref{tab-parametros} lists the number of parameters for each architecture. Both ResNet50 and DenseNet-121 are widely used for feature extraction tasks and have been tested and validated in various studies. On the other hand, MobileNetV2 offers a lightweight configuration, making it computationally less demanding. This is an important feature when running the application on computers with limited processing capabilities.

\begin{table}[ht]
\centering
\begin{tabular}{cc}
    \hline Architecture & Parameters (Millions) \\
    \hline ResNet50 & 25.56 \\
     DenseNet-121 & 7.97 \\
     MobileNetV2 & 3.4 \\
     \hline
\end{tabular}
\caption{Number of parameters (millions) in each model.}
\label{tab-parametros}
\end{table}

In both experiments, two training phases were conducted, each consisting of 50 epochs, a batch size of 32, and the Adam optimizer \citep{kingma2017adam} with an initial learning rate of 0.0001, which decreased by 30\% (gamma rate) per step, along with a weight decay of 0.0004. The MarginLoss \citep{margin} was used as the loss function. All three architectures used in this work were already imported and pre-trained using the ImageNet dataset \citep{Deng2009}. First, the conventional training of the convolutional neural network was performed using the P-NDB-UFES dataset, saving its weights and biases at the end of training. In the second stage, the NDB-UFES dataset containing the original images with dimensions of 2048x1536 was utilized. Both the NDB-UFES and P-NDB-UFES datasets were divided as follows: 5/6 for training and 1/6 for testing, with images resized to 224x224 pixels.

For evaluation, the \textit{Top-k} accuracy was used, which is defined as follows:

\begin{equation}
\text {Top-} k=\frac{\sum_{\mathbf{x}_i \in \mathcal{X}_{\text {test }}} \mathbb{I}\left(\exists \mathbf{x}_i \in \mathcal{F} \text { s.t. } y_i=y_q\right)}{\mid \mathcal{X}_{\text {test } \mid}},  \mathcal{F}=\underset{|\mathcal{F}|=k}{\operatorname{argmin}} \,\, d \left(\phi\left(\mathbf{x}_i\right), \phi\left(\mathbf{x}_q\right)\right)
\end{equation}

\noindent The \textit{Top-k} accuracy counts as correct when the class of the input query image is among the top $k$ most similar images returned by the proposed method. When $k$ is one, this task becomes a classification one, where the class of the returned image needs to be the same as the input image. In this paper, \textit{Top-1} and \textit{Top-5} accuracies are computed.

%In terms of time measurement, three time metrics were recorded: the time spent on the model, the time spent on the search, and the total time. In this work, \textit{Top-1} and \textit{Top-5} accuracy are computed.
%

The experiments were conducted using an Intel i9-7900X CPU (20 cores) with a clock speed of 4.3GHz, equipped with 128 GB of RAM, and an NVIDIA TITAN Xp graphics card with 12GB of memory. The entire code was written in Python, using various open-source libraries such as PyTorch, NumPy, Matplotlib, and Faiss.

\subsection{Results and Discussions}
\label{sec-projeto-resultado}

Next, the results obtained are presented and discussed.

\subsubsection{Experiment I - CBIR using only Images}
\label{sub-sec-projeto-resultado-experimento1}
In Experiment I, only the features extracted from the histopathological images were used. Table~\ref{tab-projeto-resultados-experimento-1-l2} presents the accuracy obtained for each CNN.

\begin{table}[ht]
\centering
\begin{tabular}{ccc}
\hline
\multicolumn{3}{c}{Experiment I - Results} \\
\hline
\textbf{Architecture} & \textbf{\textit{Top-1} (\%)} & \textbf{\textit{Top-5} (\%)} \\
\hline
ResNet50 & \textbf{74.36} & 87.18 \\
MobileNetV2 & 56.41 & \textbf{89.74} \\
DenseNet-121 & 48.72 & 84.62 \\
\hline
\end{tabular}
\caption{Results obtained using only the images.}
\label{tab-projeto-resultados-experimento-1-l2}
\end{table}

\begin{figure}[!htb]
\centering
\includegraphics[width=.8\textwidth]{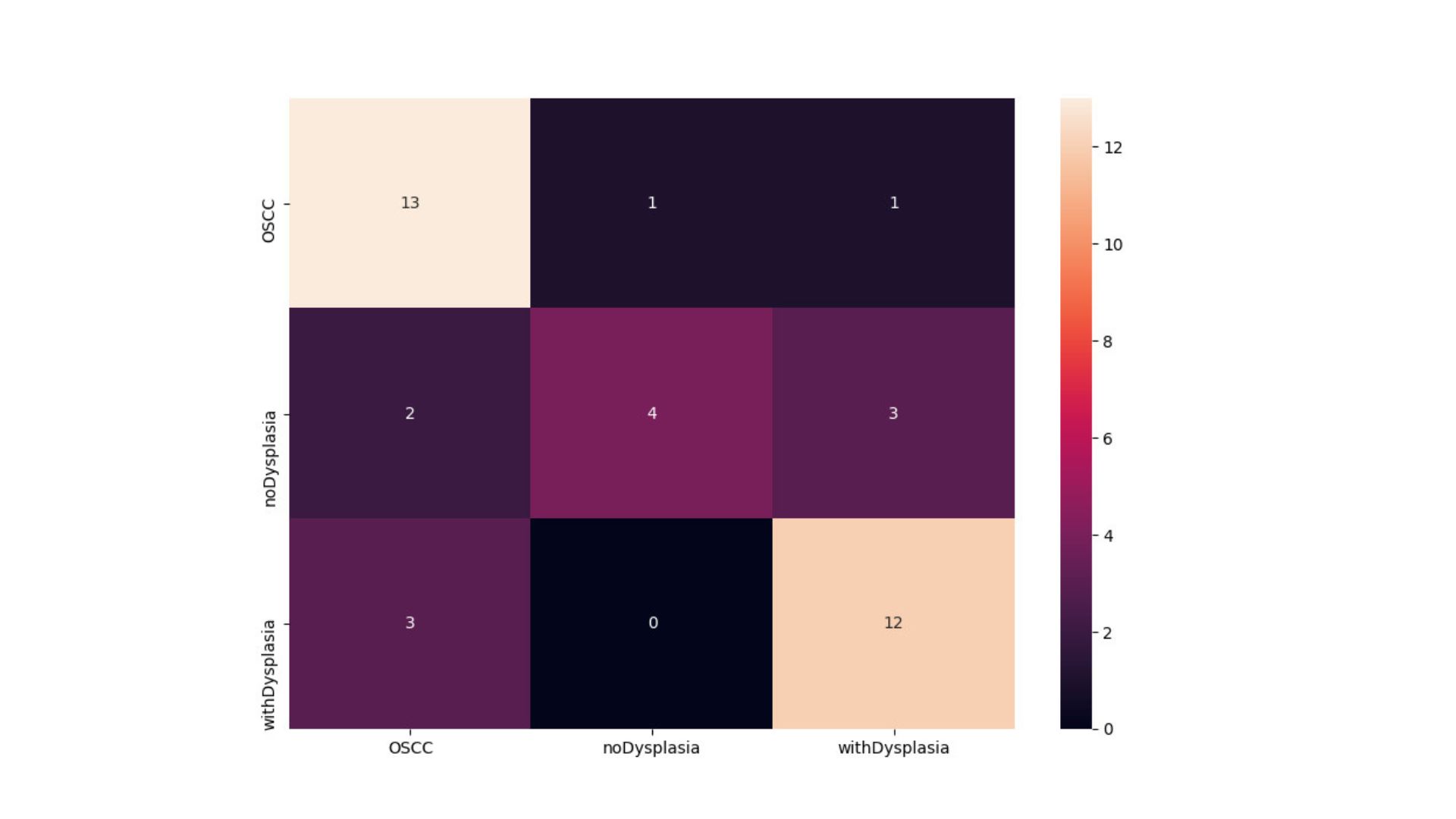} 
\caption{Confusion matrix obtained by ResNet50 in Experiment I.}
\label{fig-experimento1-matriz}
\end{figure}

By analyzing Table~\ref{tab-projeto-resultados-experimento-1-l2}, it can be observed that ResNet50 achieved a \textit{Top-1} accuracy of 74.36\%, which is the best result. The MobileNetV2 and DenseNet-121 architectures achieved Top-1 accuracies of 56.41\% and 48.72\%, respectively. In terms of the \textit{Top-5}, MobileNetV2 achieved 89.74\%, followed by ResNet50 with 87.18\% and DenseNet-121 with 84.62\%. The confusion matrix obtained in Experiment I for the ResNet50 architecture, which overall achieved the best results for \textit{Top-1}, is presented in Figure ~\ref{fig-experimento1-matriz}. The significant improvement from \textit{Top-1} to  \textit{Top-5} accuracy is expected since only one correct image among the top 5 is required to be considered correct by the \textit{Top-5} metric.

\begin{figure}[!ht]
\centering
\includegraphics[width=1\textwidth]{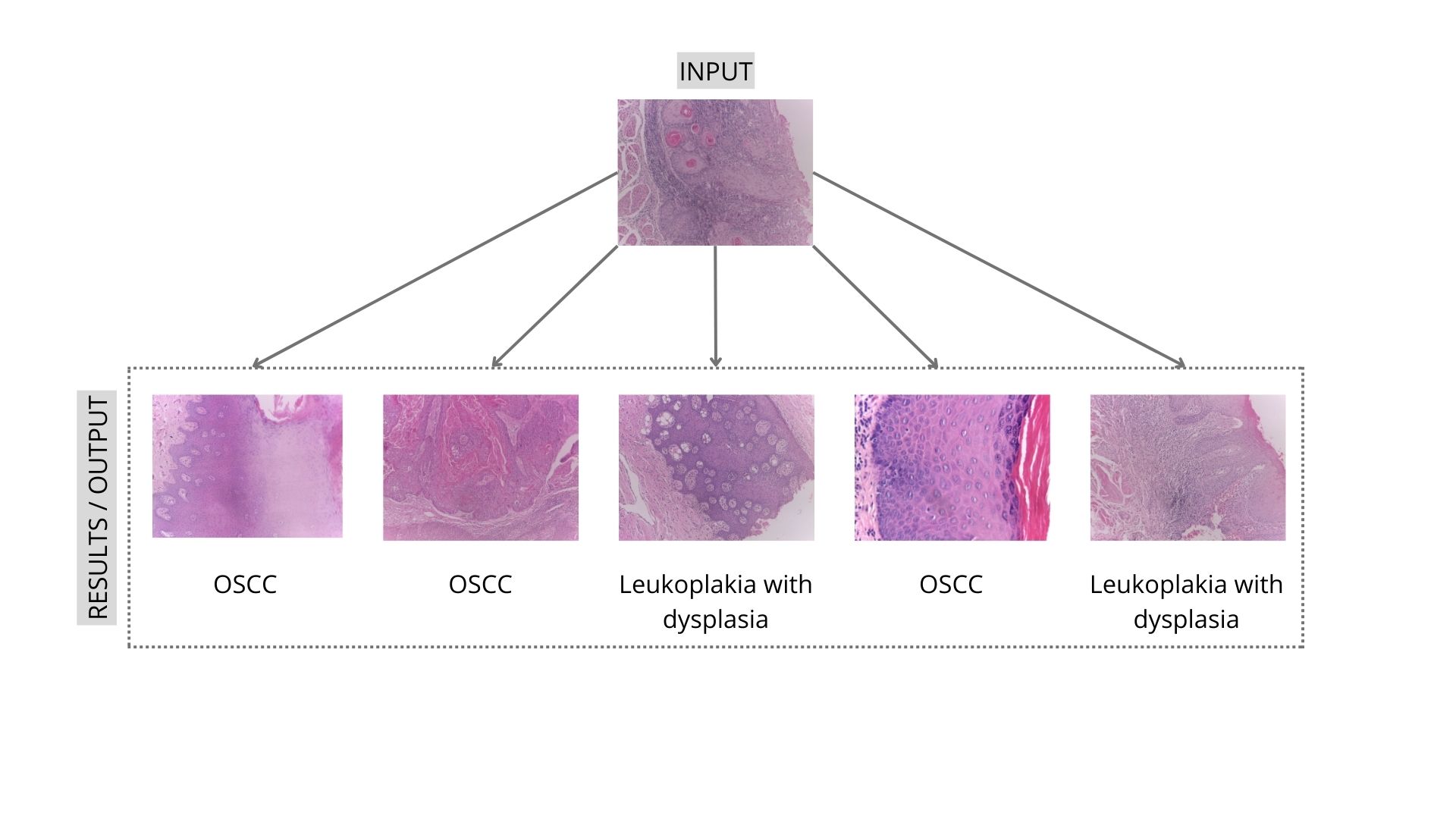} 
\caption{Example of response when receiving an input query image. The first image in the figure illustrates the input query image (above), while the others (below) are arranged as the most similar to the input query image. In this test, the query image was from the OSCC class, and the most similar ones were: OSCC, OSCC, leukoplakia with dysplasia, OSCC, and leukoplakia with dysplasia.}
\label{fig-teste-experimento1}
\end{figure}

Next, we show in Figure~\ref{fig-teste-experimento1} the 5 most similar images to an input query image using the ResNet50 configuration that achieved a \textit{Top-5} accuracy of 87.18\%. In this test, the input query image was previously labeled by the pathologist as OSCC (oral squamous cell carcinoma), and among the returned images, 3 were from the OSCC class, with the top two ranked as the best matches, and two images from the leukoplakia with dysplasia class.

\subsubsection{Experiment II - Results obtained with the proposed method CBIDR}
\label{sub-sec-projeto-resultado-experimento2}

In Experiment II, clinical data of patient was used in the image retrieval process. The TOPSIS algorithm considers two criteria: the $L_2$ between the feature vector extracted from the input query image with those stored in the Faiss library \citep{Faiss}, and the Hamming distance between the corresponding clinical data of the query image with those saved in the database. Therefore, it is necessary to specify the weights that each criterion assume in the final decision, with values between 0 and 1. The following weight sequence was used, respecting the minimum value of 0.5 for the distances obtained by Faiss: $W = [0.5, 0.5], [0.6, 0.4], [0.7, 0.3], [0.8, 0.2], [0.9, 0.1]$. The weight combination [1.0, 0.0] would be equivalent to Experiment I and was therefore not considered.

\begin{figure}[!htb]
\centering
\includegraphics[width=0.8\textwidth]{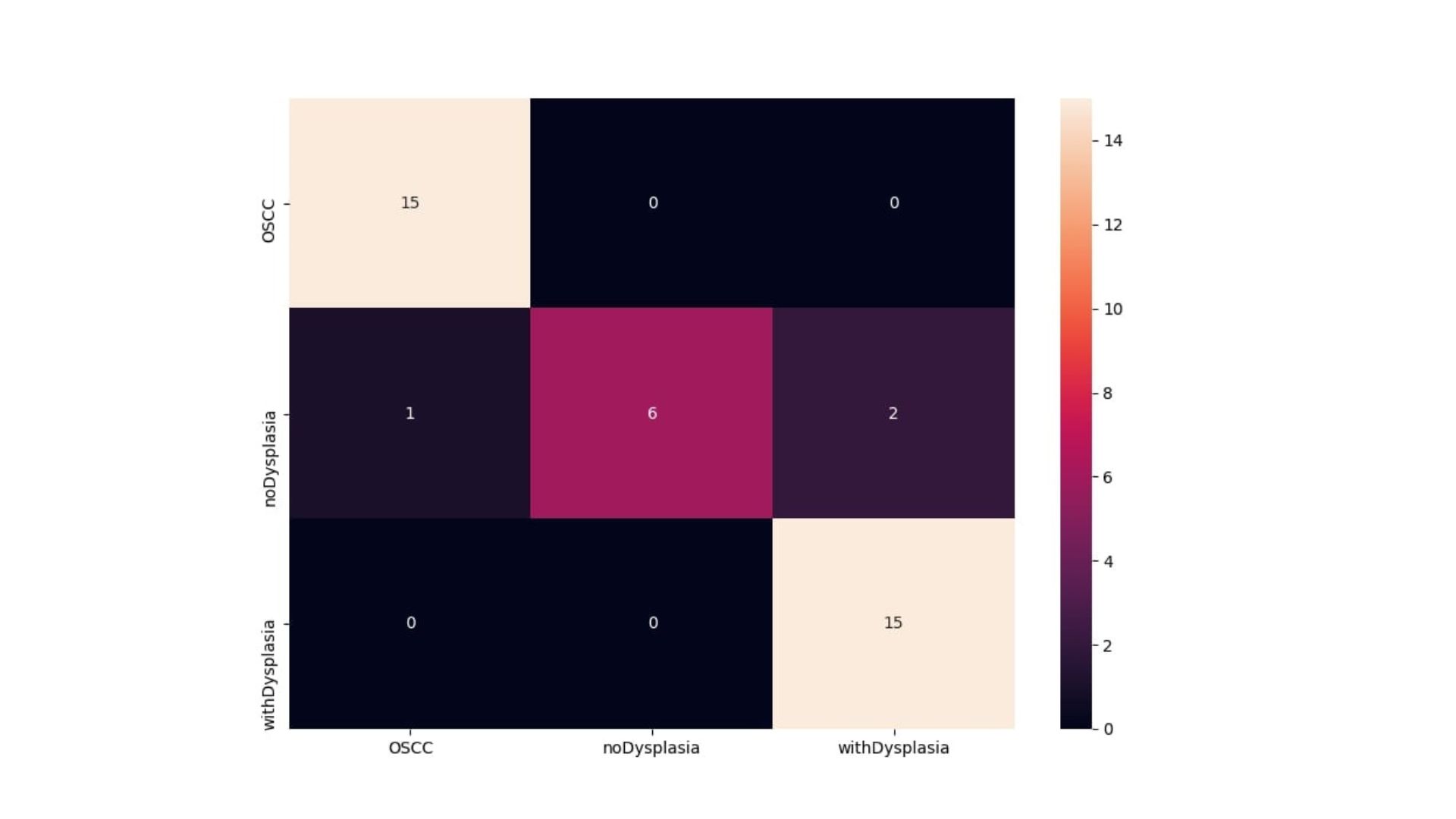} 
\caption{Confusion matrix obtained by MobileNetV2 with weights $[0.5, 0.5]$ in Experiment II.}
\label{fig-experimento2-matriz}
\end{figure}

Table~\ref{tab-projeto-resultados-experimento-2-l2} presents the accuracy for each CNN. One notes in Table~\ref{tab-projeto-resultados-experimento-2-l2}, that the best result was achieved by MobileNetV2, obtaining 97.44\% for \textit{Top-1} and 100\% for \textit{Top-5} accuracy with the weights [0.5, 0.5], indicating equal importance for both sources of information. It is worth noting that the ResNet50 and DenseNet-121 architectures also achieved their best results using the weights [0.5, 0.5]. Figure ~\ref{fig-experimento2-matriz} depicts the confusion matrix obtained using the MobileNetV2 architecture with weights [0.5, 0.5]. 

%The MobileNetV2 with the best configuration had a total time of 8.19 seconds whereas 0.59 seconds corresponded to time spent to model, and 0.62 seconds to search time. 
%

%Devido aos baixos resultados apresentados pela similaridade cosseno, eles foram listados no apêndice~\ref{app-experimento2}, assim como as matrizes de confusão e as métricas de tempo de cada arquitetura.

\begin{table}[!htb]
\centering
\begin{tabular}{cccc}
\hline
\multicolumn{4}{c}{Experiment II - Results} \\
\hline
\textbf{Architecture} & \textbf{\textit{Top-1 (\%)}} & \textbf{\textit{Top-5 (\%)}} & \textbf{TOPSIS} \\
\hline
ResNet50 & 92.31 & \textbf{100.00} & [0.5, 0.5] \\
ResNet50 & 89.74 & 94.87 & [0.6, 0.4] \\
ResNet50 & 84.62 & 92.31 & [0.7, 0.3] \\
ResNet50 & 79.49 & 92.31 & [0.8, 0.2] \\
ResNet50 & 76.92 & 87.18 & [0.9, 0.1] \\
MobileNetV2 & \textbf{97.44} & \textbf{100.00} & [0.5, 0.5] \\
MobileNetV2 & 89.74 & 97.44 & [0.6, 0.4] \\
MobileNetV2 & 84.62 & 94.87 & [0.7, 0.3] \\
MobileNetV2 & 84.62 & 94.87 & [0.8, 0.2] \\
MobileNetV2 & 64.10 & 97.44 & [0.9, 0.1] \\
DenseNet-121 & 89.74 & \textbf{100.00} & [0.5, 0.5] \\
DenseNet-121 & 82.05 & 92.31 & [0.6, 0.4] \\
DenseNet-121 & 79.49 & 94.87 & [0.7, 0.3] \\
DenseNet-121 & 74.36 & 92.31 & [0.8, 0.2] \\
DenseNet-121 & 71.79 & 92.31 & [0.9, 0.1] \\
\hline
\end{tabular}
\caption{Results obtained using images and clinical data.}
\label{tab-projeto-resultados-experimento-2-l2}
\end{table}

\begin{figure}[!htb]
\centering
\includegraphics[width=1\textwidth]{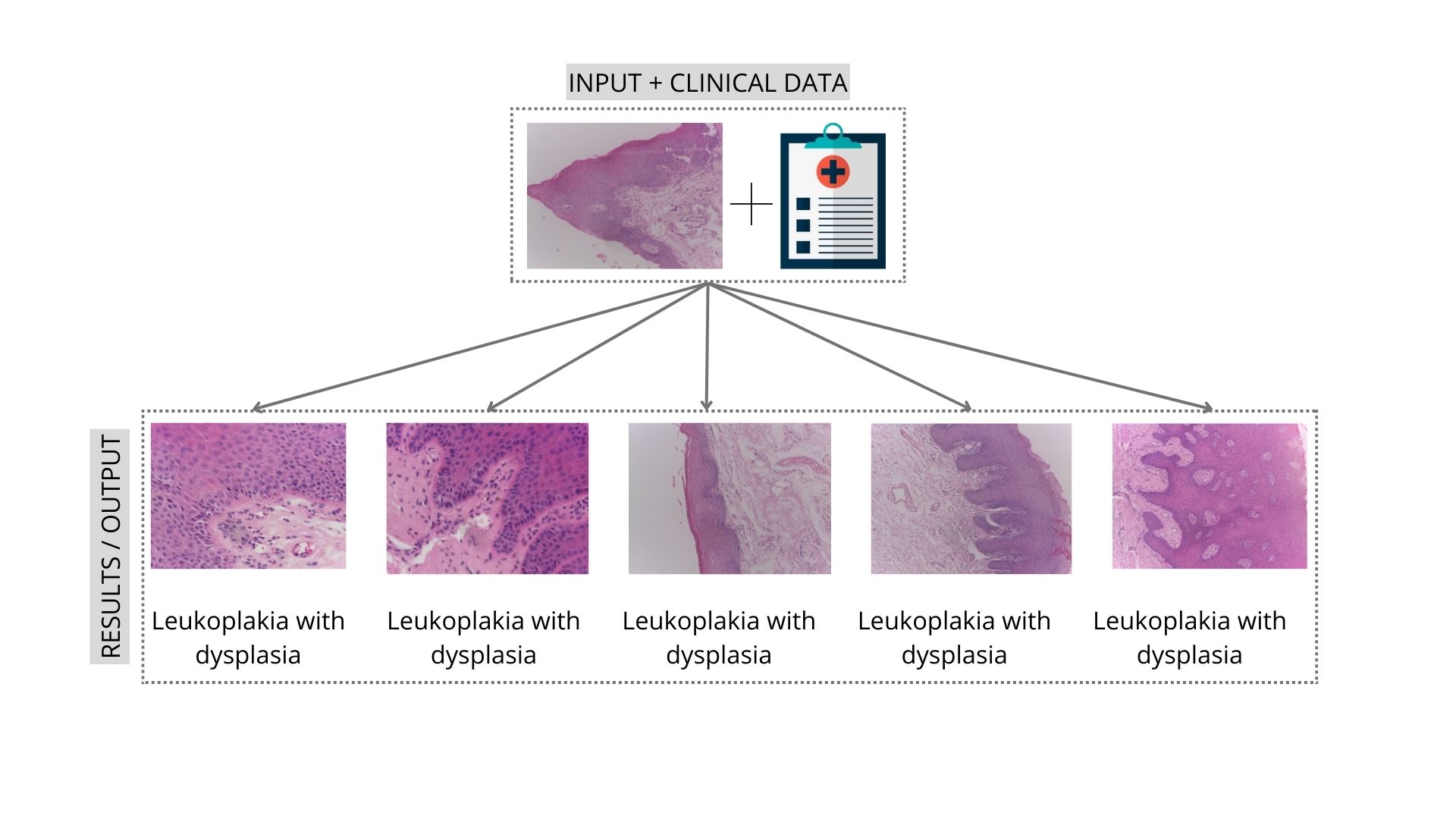} 
\caption{The first image in the figure illustrates the input query image, while the others are arranged as the most similar to the input image. In this test, the query image was from the leukoplakia with dysplasia class, and all the returned alternatives also belong to this class.}
\label{fig-teste-experimento2}
\end{figure}

We show also in Figure~\ref{fig-teste-experimento2} the 5 most similar images to an input query image and their respective clinical data, using the MobileNetV2 architecture with weights [0.5, 0.5], which achieved 100\% \textit{Top-5} accuracy. It is worth mentioning that in this test, the input query image was previously labeled by the pathologist as leukoplakia with dysplasia, and among the returned images, all 5 were also from the class leukoplakia with dysplasia.

\subsubsection{Discussions}

\label{sub-sec-projeto-consideracoes}

In general, a significant improvement in terms of accuracy is observed by using the proposed approach combining images and clinical data of patient. For \textit{Top-1} accuracy, the best configuration was the ResNet50 with value of 74.36\%, which was surpassed by 23.08\% by MobileNetV2 and TOPSIS with weights [0.5, 0.5]. In terms of \textit{Top-5} accuracy, all three models achieved 100\% when using clinical data, surpassing the first approach, which provided a maximum valaue for accuracy of 89.74\% with MobileNetV2.

\cite{Belga}, achieved 82\% and 94\%, respectively, in the \textit{Top-1} and \textit{Top-5} metrics using MarginLoss as the loss function on the the TCGA database. \cite{deLima2023} achieved 83\% balanced accuracy when using ResNetV2 to classify images from the NDB-UFES database along with clinical patient data. In both cases, as we know, it is not possible to directly compare the results obtained in this work with the ones mentioned above. The experiment II indicates promising results with the proposed method CBIDR for histopathological image combined with clinical data of patient.

\section{Conclusion}
\label{sec-conclusoes}

This paper proposed a novel method to support medical professionals in their diagnostic tasks. The methodology combines feature extraction from images using convolutional neural networks with the clinical data obtained from patients during their consultations. We illustrate the method by using the NDB-UFES dataset collected and curated by pathologists to diagnostic of oral cancer. Two experiments were conducted. In Experiment I using standard CBIR considered only the features extracted from the images as baseline  method, the ResNet50 achieved the \textit{Top-1} accuracy, with value of 74.36\% and the MobileNetV2 a value of 89.74\% for the \textit{Top-5} accuracy. In Experiment II using the proposed method CBIDR, a significant increase in accuracy  was observed, reaching 97.44\% in \textit{Top-1} and 100\% in \textit{Top-5} using  MobileNetV2. The results obtained from the  Experiment II, show that the proposed method CBIDR improved accuracy in retrieving histopathological images with clinical data of patients. This work opens up a new research avenue for multimodal information retrieval using several information sources in an easy and effective way. For future work, we aim to make the approach usable to medical professionals (doctors/pathologists) by creating a friendly graphical interface to allow for more intuitive use. Additionally, we also intend to apply the proposed approach to other medical datasets.

\section{Acknowledgments}
R.A. Krohling thanks the Brazilian research agency Conselho Nacional de Desenvolvimento Científico e Tecnológico (CNPq), Brazil - grant no. $304688/2021-5$ and the Fundação de Amparo à Pesquisa e Inovação do Espírito Santo (FAPES), Brazil – grant no. $21/2022$.

\section{Declaration of Competing Interest}
The authors declare that they have no known competing financial interests or personal relationships that could have appeared to influence the work reported in this paper.

%\section{Ethics approval and consent to participate}
%Not Applicable.

\section{Role of the Funding Source}
The funders had no role in study design, data collection and analysis, decision to publish, or preparation of the manuscript.

\section{CRediT authorship contribution statement}
\textbf{Renato Antonio Krohling:} Conceptualization, Methodologies, Investigation, Validation, Writing - Review \& Editing, Supervision.
\textbf{Humberto Giuri Calente:} Software, Validation, Writing - Original Draft, Review \& Editing.


%\bibliographystyle{elsarticle-num} 
\bibliographystyle{apalike} 
\bibliography{bib-abbrev}

% \bibliographystyle{elsarticle-num-names}

% \bibliography{bib}
%\bibliography{bib-abbrev}





% \end{thebibliography}
\end{document}